# Optical Two-Tone Time Transfer

Jonathan D. Roslund*, Abijith S. Kowligy, Junichiro Fujita, Micah P. Ledbetter, Akash V. Rakholia, Martin M. Boyd, Jamil R. Abo-Shaeer, Arman Cingöz*

Vector Atomic, Inc., Pleasanton, CA, USA

## Abstract

Sub-picosecond timing synchronization can enable future optical timekeeping networks [1], including coherent phased array radar imaging at GHz levels [2], intercontinental clock comparisons for the redefinition of the second [3], chronometric leveling [4], and synchronization of remote assets, including future satellite-based optical time standards. With optical clocks now operating on mobile platforms [5], free-space synchronization networks with compatible performance and the ability to operate under platform motion are essential to expand the reach of precision timing. Recently, femtosecond (fs)-level optical time-transfer techniques have been developed that can operate over hundreds of kilometers despite atmospheric turbulence, signal fade, and dropouts [6] [7] [8]. Here we report a two-tone optical time transfer scheme with comparable performance that reduces hardware requirements and can support both fiber and free-space networks. Using this technique, sub-fs synchronization was demonstrated over a ~100 m free-space link for several hours. In addition, the link was used to syntonize two iodine optical clocks and then compare them over four days. The set-up employs an integrated photonics transceiver and telecom-band lasers that are compatible with full photonic integration.

## Introduction

With the development of optical atomic clocks, advances in timing precision have greatly outpaced global-scale synchronization capabilities. Laboratory optical clocks now achieve sub-femtosecond timing jitters at 1 second and fractional frequency instabilities at $10^{-18}$-levels [9]. Current free-space microwave timing links (e.g., TWSTFT, GPS carrier phase [10] [11] [12] [13], ACES-MWL [14]) achieve instabilities of 10-100 picoseconds (ps); optical laser ranging (i.e., T2L2 [15]) realizes similar performance.

To address this performance gap, novel optical time-transfer techniques have been developed. In one approach, stable trains of ultrashort optical pulses are exchanged over a free-space link to provide femtosecond-level synchronization [16] [17] at distances over 300 km with only picowatts of received power [6]. Along with the primary frequency comb, which is a fundamental component of the optical reference, this method employ a second frequency comb at each network node, adding complexity, power consumption, and expense to the optical transceiver. In addition, the transceiver requires careful dispersion management for the transmission of ultrashort pulses. An alternative approach uses a continuous-wave (CW) laser to transfer timing information between remote sites, analogous to a technique originally developed for cancelling Doppler shifts in fiber links [18] [19]. This scheme shifts system complexity from the laser system to the optical terminal, since adaptive optics are necessary to track the optical carrier phase without cycle slips over turbulence [20].

* Corresponding authors: jon@vectoratomic.com, arman@vectoratomic.com

Here we report the development of a simplified optical time transfer scheme that reduces the size, weight, power, and cost (SWaP-C) of the overall system while maintaining sub-femtosecond synchronization at 1 second with a precision below $10^{-17}$. This novel two-way time transfer technique replaces the second frequency comb in Ref [16] with two telecom distributed-feedback (DFB) lasers. The two DFB lasers are transmitted across the link, and the differential optical phase experienced between them measures the optical group delay associated with the link. This two-tone approach measures the same information encoded in the temporal pulse but does so with a frequency domain projective technique. Related approaches have been employed for precision ranging and frequency transfer measurements across free-space optical links [21] [22] [23] [24].

The two DFB lasers provide operational simplicity and reduced SWaP-C. DFB lasers are also amenable to chip-scale integration for additional future SWaP-C savings (see Discussion). Additionally, the use of CW lasers, particularly telecommunication lasers designed to operate on the fixed international telecommunication union (ITU) grid, allows for an optical transceiver that is directly compatible with existing coherent optical terminals (e.g. [25]). As discussed below, key performance metrics, such as resolution and dynamic range, are easily adjusted to optimize system performance for particular applications. Finally, removing the requirement to track timescale variations at optical carrier levels eliminates the need for adaptive optics at the terminals.

The timing information necessary for synchronization is derived from an optical transceiver contained within a simplified, purpose-designed photonic integrated circuit (PIC), which minimizes out-of-loop paths between the timescale reference and measurement planes. The result is sub-femtosecond temporal synchronization for time periods exceeding $10^4$ s, equivalent to a frequency resolution $< 10^{-19}$. This performance is sufficient to synchronize state-of-the-art optical clocks for extended durations with reasonable system complexity and size. Finally, we present a practical demonstration of the technique that continuously compares two iodine optical clocks [5] over a free-space link for four days.

## Description of the Technique

The two-tone technique is based on the principle that any variation in the envelope arrival time for a pulse train is equivalently captured in the frequency domain where the comb spectrum exhibits a linear spectral phase with a slope given by the arrival time (Figure 1) [26]. The linear component of this spectral phase and thus any deviation of the underlying timescale can be reconstructed by sampling the spectral phase at only two frequencies.

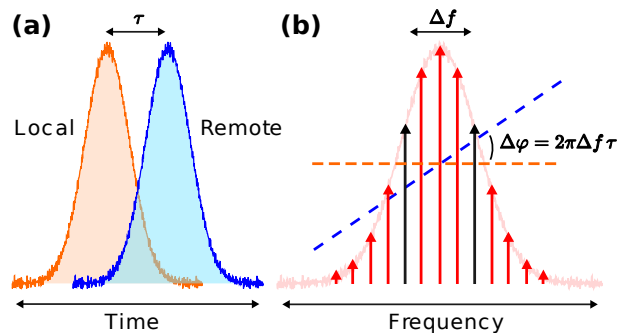

Figure 1. Two-tone time transfer concept. (a) Optical pulse delay between local (orange) and remote (blue) pulses. The difference in their respective timebases is τ; (b) Frequency domain representation wherein the delay τ creates a linear spectral phase Δφ=2πfτ with respect to the reference pulse. The delay τ is extracted by sampling this phase at two points separated by Δf.

Consider a timing network with two sites, A and B, that have a free-space or fiber optic link between them (Figure 2). At site A, two continuous-wave (CW) lasers with frequencies $f_A^{(1),(2)}$ and frequency spacing $\Delta f = f_A^{(1)} - f_A^{(2)}$ are each phase-locked to the frequency comb of an optical reference. The self-referenced frequency comb co-located with the reference operates as a transfer oscillator, imprinting the comb spectral phase at the two CW laser frequencies onto their respective carrier phases. Frequency instability of the clock becomes a variation of the envelope arrival time for each comb pulse with respect to a local reference plane. This arrival time variation is then manifest as a differential optical phase between the two CW lasers. The frequency reference at Site B contains its own frequency comb and two CW lasers $f_B^{(1),(2)}$ with the same frequency spacing $\Delta f = f_B^{(1)} - f_B^{(2)}$. These two lasers are similarly phase-locked to the site B comb.

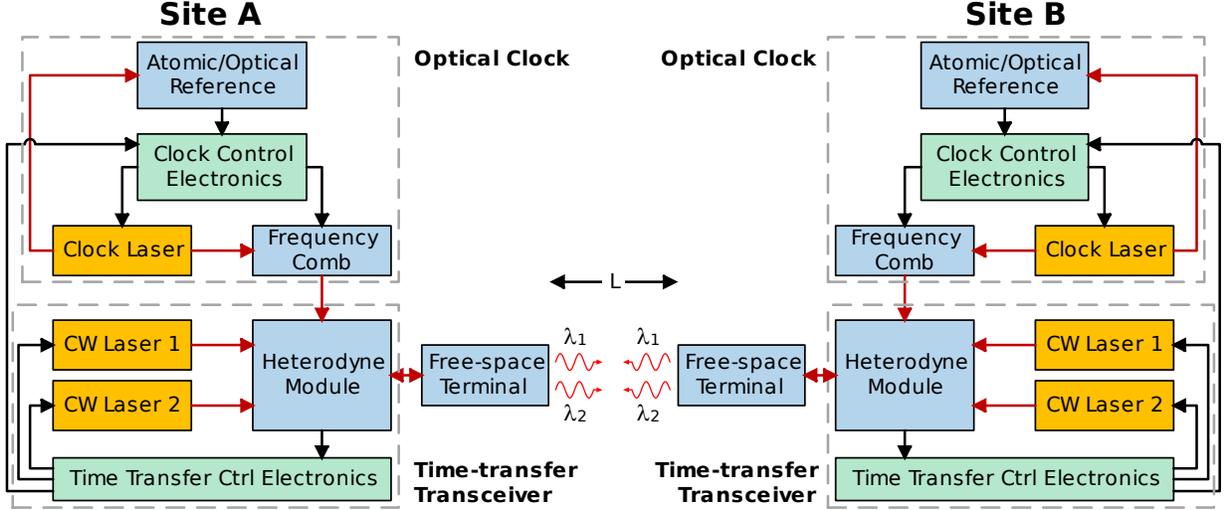

**Figure 2.** Optical two-tone time transfer scheme. Each site contains an optical reference to which a frequency comb is stabilized. Light from the frequency comb is introduced onto a heterodyne transceiver along with two CW lasers. These CW lasers are phase stabilized to the comb, which maps the temporal jitter of the comb pulses onto their differential optical phase. The CW lasers are exchanged between the two sites, and heterodyne beats between the local and remotely received lasers reveal the one-way group delay, which includes the link time-of-flight as well as timing error between the two reference oscillators. One CW laser at each site is utilized for classical communication to exchange its one-way timing information, which allows deducing the two-way relative timebase error between the sites.

Both pairs of CW lasers are then exchanged across the optical link. An optical transceiver at each site separates the two incoming CW lasers and optically interferes them with the local copies. A heterodyne beat for each CW pair is detected at the difference of the site A and B offset lock frequencies, and demodulation of this beat reveals the optical phase accumulated between the two sites, $\varphi_{A \to B}^{(k)} = 2\pi f^{(k)} \tau_{A \to B}^p$, where $k = 1$ or 2 and $\tau_{A \to B}^p$ is the phase delay across the link. The one-way group delay across the link $\tau_{A \to B}^g$ is then recovered from the difference in the optical phases accumulated by the two lasers, $\delta\varphi_{A \to B} = \varphi_{A \to B}^{(1)} - \varphi_{A \to B}^{(2)} = 2\pi \Delta f \tau_{A \to B}^g$. The group delay $\tau_{A \to B}^g = T_{A \to B} + \Delta t_{A,B}$ contains both the transit time $T_{A \to B}$ between the two separated temporal reference planes as well as the relative timescale difference $\Delta t_{A,B}$. The same processing scheme occurs at the second site to recover $\tau_{B \to A}^g = T_{B \to A} - \Delta t_{A,B}$, and the individual group delays measured at each site are combined to separate relative clock drift from transit time variations. Assuming a reciprocal optical link ($T_{A \to B} =$ $T_{B\to A}$), the relative difference in the timescales at site A and B is $\Delta t_{A,B} = (\delta\varphi_{A\to B} - \delta\varphi_{B\to A})/4\pi\Delta f$. Thus, the two-tone detection scheme is equivalent to directly measuring a temporal envelope shift for pulses of bandwidth $\Delta f$. The tradeoff between sensitivity and dynamic range is determined by the spacing $\Delta f$ between the two CW lasers, which is adjustable depending upon the application. The minimum timing deviation measurable with the two-tone technique is:

$$\Delta t_{A,B,min} = \frac{T}{2\pi} \times \sqrt{2/N + \kappa \times \Delta_{rms}^2}$$

where $T = 1/\Delta f$ is the period of the synthetic fringe, $\Delta_{rms}$ is the RMS phase noise for one of the CW optical phase locks to the comb, $N$ is the total number of signal photons detected for both sites per acquisition period, and $4 \leq \kappa \leq 8$ is a numerical constant that depends upon the link length (see Supplemental Material).

Using two CW lasers for generating the continuous synthetic fringe signal makes efficient use of all detected photons over the link. In addition, since the typical synthetic fringe period is ~1,000× larger than the individual optical cycles of each CW laser (5 fs period at 1550 nm), the technique is tolerant to signal dropouts from turbulence or weather [27]. Signal dropouts do not affect the timescale continuity if the two clocks do not drift away from each other by more than the period of one synthetic fringe during the dropout. For a CW laser separation of 100 GHz, corresponding to a single ITU grid channel spacing, the synthetic fringe has a period of 10 ps. State-of-the-art optical clocks maintain picosecond-level holdover for many days [28] and portable optical clocks maintain it for several hours [5]. Thus, cycle slips in the observed timescales would not occur even for extended dropouts, e.g., low earth orbit periods (~1.5 hours).

Dynamic range beyond one synthetic fringe is readily extended using a variety of techniques. For the comb-based time transfer technique, a separate CW laser with an electro-optic phase modulator (EOM) at each site was used to establish a coarse time transfer system [17], which can be easily incorporated into the two-tone technique. An alternative approach utilizes a phase modulator to create a static sideband. The frequency spacing between the carrier and the sideband defines a second synthetic wavelength. The modulation frequency is chosen to resolve the picosecond (millimeter) range ambiguity of the CW laser synthetic wavelength [29] [30]. For example, 100 MHz phase-modulation creates a synthetic wave with period 5 ns, sufficient to resolve the uncertainty of GPS synchronization. Finally, the sideband frequency for coarse timing is tunable in real time and can be adapted to optimize the system for different applications.

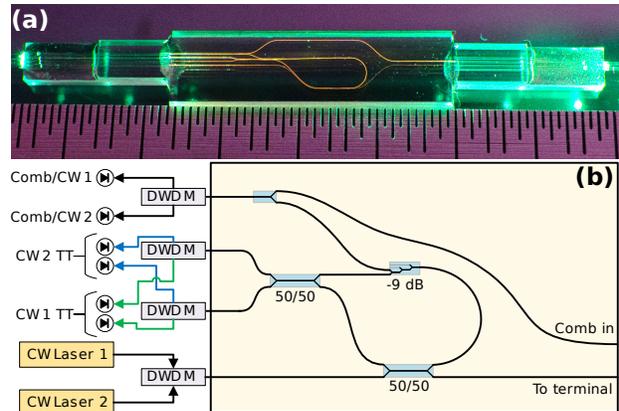

Figure 3. PIC transceiver for two-tone time transfer. a) Photograph of the fiberized transceiver. Green light is coupled onto the PIC to illuminate the waveguides. For scale, the finest gradation on the adjacent ruler is 0.5 mm. b) Schematic of the PIC transceiver. Overlap of all relevant fields occurs on-chip, which minimizes out-of-loop pathways to the millimeter scale.

## Demonstration

Our initial demonstration uses two 1550 nm DFB lasers separated by two DWDM channels on the ITU grid ($\Delta f = 200$ GHz), providing a synthetic fringe period of 5 ps. All optical multiplexing/de-multiplexing is done with discrete telecom components (DWDM filters) at the input and output of the PIC, which combine or separate the light in the two ITU channels. The PIC (Figure 3a) consists of a 6-port device fabricated on a 10 mm × 20 mm glass substrate that uses an ion exchange process to define the waveguides directly on glass. V-grooves are attached to the substrate ends for fiber coupling the inputs and outputs; excess loss at the PIC fiber interfaces is < 0.5 dB per facet. The PIC design (Figure 3b) includes a 50/50 coupler that sends half the power for each local laser to the free-space terminal port. The other half is attenuated by 10 dB and interfered with the optical signals received from the remote site (over the same terminal port) using a second coupler. A portion of the light dumped at the attenuator is combined with the comb light to generate the beatnotes necessary to phase-stabilize the CW lasers to the frequency comb. This topology eliminates any differential path length between the two local CW lasers and minimizes out-of-loop pathways between the comb stabilization reference plane and that for the interference with remote light.

Each site uses a custom electronic circuit for balanced photodetection and signal conditioning, which delivers amplified signals to a FPGA. The FPGA phase-stabilizes the two CW lasers to a 200 MHz fiber frequency comb and analyzes the beatnotes between local and remote CW lasers to retrieve the one-way group delay.

## Common Clock Demonstration

The sensitivity limit for the two-tone technique was assessed by co-locating both sites and stabilizing all four CW lasers to the same frequency comb. This was accomplished by dividing the common frequency comb between the two PICs with a telecom 50/50 splitter, which minimizes relative clock drift between the sites. Any remaining clock noise is attributed to either differential path lengths between the comb split and the two PICs or the technique itself. Following phase-stabilization of the two CW lasers on the PIC, the output from each PIC was transferred to a breadboard free-space link with a 5-meter optical fiber and exited through an optical collimator. The two collimators were separated by ~1 m, with one collimator mounted on a translation stage to precisely adjust the link distance. One-way group delays were acquired at each site while adjusting the translation stage. The sum and difference of the one-way delays yield the link time-of-flight and the residual timing difference between the two sites, respectively (Figure 4a). The time-of-flight signal closely mirrors the individual one-way group delays measured at each site, and a translation of ~30 ps, corresponding to ~6,500 optical fringes, is tracked without any wraps. The timing signal, however, maintains femtosecond instability (sub-fringe) throughout the translation due to the reciprocal link. The data processing algorithm to track one-way group delays larger than the ambiguity range of the system is discussed in Methods.

The timing deviation noise floor measured as a function of received power is shown in Figure 4b. Above 100 pW, the signal-to-noise ratio (SNR) is power-independent and limited by residual technical phase noise on the four CW laser offset locks, which provides a timing sensitivity of ~400 attoseconds with 1 s averaging. Conversely, below 100 pW the deviation scales as $1/\sqrt{N}$.

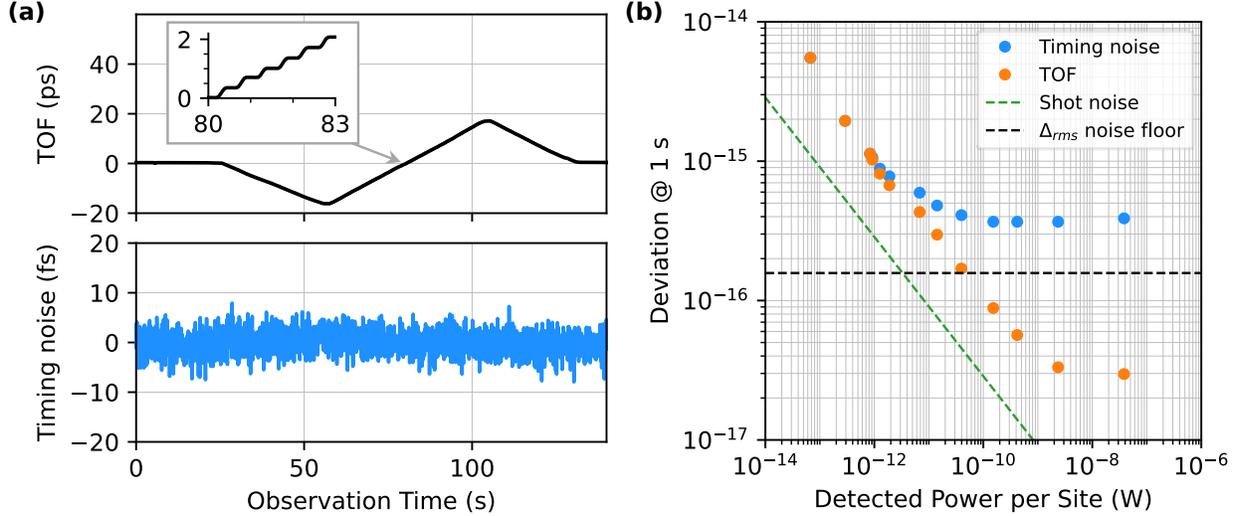

Figure 4. Demonstration of the two-tone concept and detection sensitivity. A) The distance between two sites is varied in a controlled manner to provide a temporal displacement of 33 ps (equivalent to ~6500 optical fringes). The time-of-flight signal (top panel) shows the expected temporal profile, including individual steps of the translation stage (inset). The two-way timing noise signal (bottom panel) exhibits temporal precision at the fs level (5 ms sample interval) throughout the induced translation. B) The TDEV sensitivity at an averaging interval of 1 s is shown as a function of the total received power at each site. For received powers exceeding 100 pW, the timing sensitivity is limited by the residual noise of the offset locks to ~400 as. Below 100 pW, the timing sensitivity scales as $1/\sqrt{N}$, but still is capable of synchronization to 1 fs with 1 pW of received power. For short links where $\kappa = 8$, this data is within ~2× of the prediction for $\Delta t_{A,B,min}$.

The measured deviation is ~2× larger than the theoretical prediction over the entire power range (see Methods). Nevertheless, 1 fs timing sensitivity is observed for only 1 pW of received power. A straightforward approach to lower the technical noise floor is to use narrower linewidth CW lasers in place of the DFBs. Such lasers would likely decrease the offset lock residual phase noise by 10× and lead to a corresponding increase in the timing resolution. Notably, the two-way TOF ranging precision is not limited by the same technical noise and exhibits $1/\sqrt{N}$ scaling for received powers up to ~10 nW. This is due to the correlation of the offset lock noise between the sites, which is removed by subtracting the two one-way group delays measured at each site. However, this classical correlation disappears for long links, and the detection limits for both the timing and TOF modes converge to a value $\sqrt{2}$ smaller than the residual noise limit arising from the phase locks seen in Figure 4b. (see Supplemental material).

To extend the free-space link, we constructed two co-located free-space optical terminals in our facility. Each terminal included 1550 nm telescope optics that expanded the beams to 7 mm. Light was launched free-space to a retroreflector ~45 meters away and returned to the adjacent terminal for a folded path length of ~90 meters. Fifty meters of optical fiber delivered the light from each PIC to the terminals, for a total link length of 200 meters (fiber and free-space combined). Two-way timing errors were measured with 100 pW of received power across the 90-meter free space link. Despite several picoseconds of fluctuation between the two terminals and the thermal sensitivity of the 110-m delivery fiber, the technique achieves 500 attosecond temporal resolution with 1 s of averaging and maintains sub-femtosecond timing deviation between the two sites over several hours (Figure 5a). The corresponding fractional frequency fluctuation (Figure 5b) is $10^{-15}$ after 1 s and below $10^{-18}$ at 1,000 s, compatible with state-of-the-art laboratory optical clocks [9] [28].

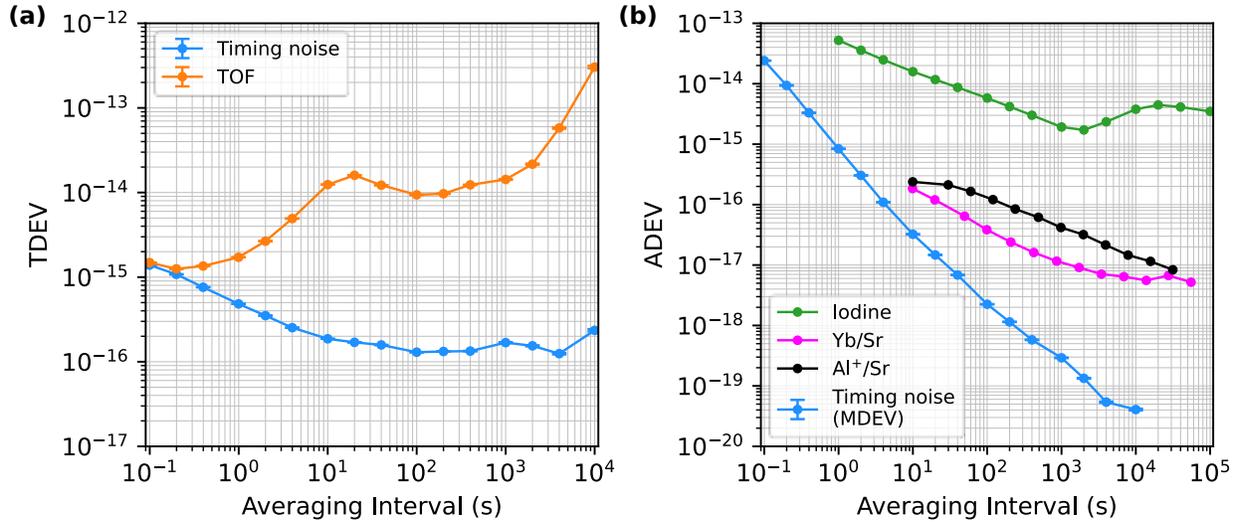

Figure 5. Common clock two-tone performance measured over 90 m free-space link with 100 pW of received power. A) The TDEV for the measured timing noise between the two sites is 500 as at 1 s of averaging and maintains < 1 fs of deviation for time periods up to $10^4$ s. Notably, the excess ~ 100 m of fiber for delivery to the optical terminals does not degrade the instability. B) The technique provides a fractional frequency instability < $10^{-18}$ at 1000 s (ADEV). The ADEV for several state-of-the-art optical clocks [9] as well as a mobile iodine optical clock [5] are shown for comparison.

Dispersion compensation was not necessary for the 110-m PM-1550 fiber delivering light to the optical terminals (dispersion equivalent to 315 km of atmosphere at sea-level). The two-tone approach has inherent immunity to dispersion, by operating in the frequency domain and the direct projection of the linear component from the entire relative spectral phase. This allows synchronization to be maintained for dynamically changing optical path lengths (e.g., satellite tracking through the atmosphere, synchronization with mobile planes or ships). A non-reciprocal timing error $\Delta t_{error} \sim 10^{-19}$ s/km of air propagation arises due the different offset frequencies at each site (see Supplemental Material).

## Clock Comparison

Following the measurements characterizing the performance of the technique using a single frequency comb common to both sites, the timing deviation between two independent iodine atomic clocks was measured over the original short free-space link. The iodine clocks exhibit short-term instabilities of ~$5\times10^{-14}/\sqrt{\tau}$ and ~$2\times10^{-14}/\sqrt{\tau}$, respectively. A portion of each clock's 200 MHz frequency comb was delivered via a 10-meter optical fiber to its corresponding transceiver PIC to stabilize the local CW lasers. The laser pairs were exchanged over the free-space link, and the resultant beatnotes were processed by the FPGA to extract the timing deviation between the clocks. A 1064-nm optical beatnote between the iodine clocks was also monitored with a conventional frequency counter for cross-validation.

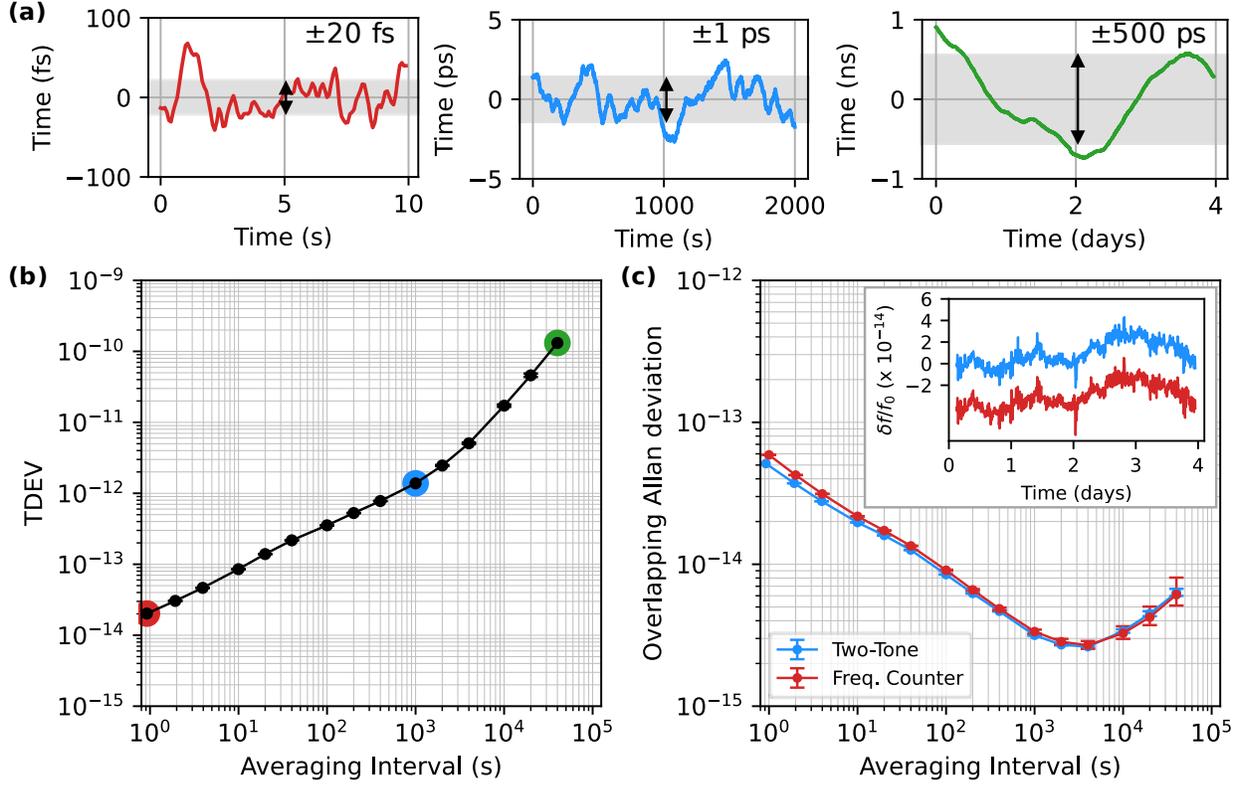

Figure 6. Comparison between two iodine optical clocks with the two-tone technique. A) Time traces highlighting the temporal instability recorded between the two clocks for several observation periods. The two-tone measurement operated continuously for a period of four days. B) Timing deviation measured over a four-day period. The shared instability between the two clocks is 20 fs at 1 s of averaging and 600 ps after one day. The colored dots correspond to the observation periods displayed in panel A. C) Overlapping Allan deviation for the fractional frequency fluctuations measured by both the two-tone technique and a frequency counter. The inset shows the raw frequency record for the two techniques over the entire four-day period.

Initially, the two clocks were deliberately frequency offset from one another by known values to verify the calibration of the two-tone technique. The observed timing deviations accumulate error linearly in time with the expected slopes (Supplemental Figure). Once the calibration was verified, one of the clock comb offset lock frequencies was updated to reduce the frequency offset to ~ $10^{-15}$.

After initial syntonization, timing fluctuations between the two clocks were measured over four days (Figure 6). A relative wander between the two timescales of ±500 ps was observed over this period, and the corresponding timing deviation is displayed in Figure 6b. The relative timing instability between the two clocks is 20 fs at 1 s of averaging (Figure 6a), approximately 40× larger than the two-tone noise floor. Moreover, these two systems maintain synchronization below 10 ps for nearly 2 hours (Figure 6b). The timing fluctuations in Figure 6a are converted to frequency variations in Figure 6c; the simultaneous frequency counter measurement is inset. The congruence between the frequency traces and Allan deviations illustrates the fidelity of the two-tone approach for comparison of optical timescales. The discrepancy between the two Allan deviations at short times is attributed to comparing results from a Λ-frequency counter (Keysight 53100A [31]) to the two-tone method that conducts Π-averaging [32] [33].

## Discussion

For the data presented above, the individual one-way time-of-flight information was extracted in real time using the data processing algorithm presented in the Methods section, and the two-way timing signal was computed in post processing. Implementation of an optical communications channel to extract the two-way timing signal in real time as well as signal processing algorithms for real-time syntonization and synchronization of the clocks is currently under development.

The two-tone approach offers several unique advantages for operation under motion. The technique does not possess an intrinsic optical Nyquist limit, and the maximum Doppler shift resolvable is determined by the bandwidth of the photoreceiver and digital processing unit. Detection of Doppler shifts arising from terminal motion with relative speeds of 1 km/s is enabled with 10 GHz photoreceivers. Digitization and processing approaching these rates are possible with modern FPGAs [34], and the required digital signal processing has been developed for radar, coherent optical communications, and radio astronomy [35] [36] [37] [38]. Notably, the bandwidth needed to follow GHz Doppler shifts is incumbent upon the digital processor and not any optical actuators [39] [40] [41] [42].

Compatibility with fiber delivery also allows the integration of the two-tone scheme into fiber-optic networks for time and frequency dissemination. The technique is wavelength agnostic and can be employed with any pair of WDM channels in the telecommunication grid, including the O, C, and L bands. This provides an opportunity to utilize existing bidirectional, long-haul fiber-optic infrastructure for time-transfer with femtosecond precision. Such functionality can enable applications including intracontinental clock comparisons [43] as well as synchronization in a large scientific facility [44]. The technique is also compatible with the 1064 nm region under consideration for future space-based optical communication links [45] [46]. Depending on the optical reference (i.e., iodine), one of the time-transfer lasers can also serve as the optical clock CW laser, reducing the laser count for the transfer link to one.

The present implementation motivates integration of additional functionality onto the transceiver PIC. Several hybrid integration strategies are commercially available to directly interface multiple DFBs with silicon (Si) or silicon nitride (SiN) platforms, including butt-coupling or flip-chip bonding of the laser to the PIC. Furthermore, narrow linewidth CW lasers on platforms such as Si or SiN [47]

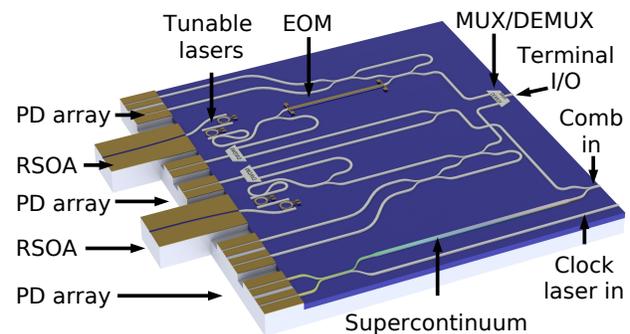

Figure 7. Photonically integrated transceiver concept: Hybrid or heterogenous integration of active and passive photonic platforms allows for integrating the entire optical transceiver, including the two CW lasers and wavelength (de)multiplexers, onto a single PIC. An on-chip EOM in one (or both) arms can be used for coarse time transfer and conventional optical communications. If the passive platform is sufficiently nonlinear, dispersion engineering allows for on-chip super-continuum generation from the frequency comb, which in turn allows for frequency comb self- and optical- referencing on the same PIC to define a single timing plane for the entire clock/time transfer system. RSOA: reflective semiconductor amplifier; PD: photodetector; EOM: electro-optical modulator; MUX/DEMUX: wavelength multiplexer/de-multiplexer.

[48] [49] [50] [51] are under development and may support the future integration of lower intrinsic linewidth lasers with the rest of the transceiver. Similarly, advances in nonlinear nanophotonics allow for supercontinuum generation and electro-optic modulation on platforms such as thin-film Lithium Niobate (LN) [52] [53] [54] [55], or hybrid wafer level integration of SiN with LN [50]. These could enable combining frequency comb referencing with the time transfer system to define a single temporal reference plane for the system. A concept model for an integrated optical transceiver is shown in Figure 7. On-chip lasers and dense wavelength division multiplexing filters can be used to multiplex and demultiplex the two wavelengths on chip. An on-chip EOM in one or both arms can be used for coarse time transfer and conventional optical communication [17] [29]. A small portion of the incoming comb light is interfered with each laser to provide the beatnote signals necessary for stabilizing the lasers to the frequency comb. The remaining comb light is used to drive octave-spanning supercontinuum generation for self-referencing. A portion of the supercontinuum is overlapped with the clock CW laser to fully stabilize the frequency comb to the optical reference.

## Conclusion

Frequency combs are a powerful tool for synchronizing optical timescales to femtosecond precision. Here we introduced a CW transfer technique that leverages the stability and benefits of the frequency comb already built into the local optical clocks in the network. This projective technique retains the benefits of comb or single laser transfer while reducing the system SWaP-C and providing new opportunities to address platform motion and operate across hybrid fiber and free space links.

Two-tone time transfer provides synchronization below one femtosecond with picowatts of received power, compatible with 100 km+ terrestrial links or ground-to-satellite links. The technique is readily extendable to multiple nodes with appropriately chosen CW laser offset frequencies. Combined with portable optical clocks, it offers a promising path to global synchronization at picosecond levels.

## Methods

### Optical Architecture

A block diagram of the optical architecture at each site is shown in Figure 2 and Figure 3b. DFB lasers centered at 1536.6 nm (ITU Channel 51) and 1535.0 nm (ITU Channel 53) are used at each site. These lasers have ~200 kHz linewidths; the 200 GHz spacing between them creates a synthetic fringe with a 5 ps period. The two lasers are multiplexed together with a telecom component and coupled into the PIC. The combined CW light is split with a 50/50 coupler, and one output is directed to the free-space terminal. The other output is attenuated and split again. A portion of this output is combined with light from a 200 MHz frequency comb, exits the chip, is demultiplexed with another telecom component, and the two CW laser-comb beatnotes are detected. The other portion is combined with half of the received light on a 50/50 coupler, after which it is demultiplexed off-chip, and each wavelength is separately detected with a balanced detector.

For this scheme, there are two non-common optical paths for which differential fluctuations manifest as noise or drift in the detected timescale. First, there is differential path between the local and received light in the heterodyne interferometer. Second, there is differential path between the local light that is phase-locked to the comb and that which optically interferes with the received light. Importantly, both differential pathways are on the PIC, which allows them to be millimeter size rather than tens of centimeters as with a fiber-based transceiver. All the fiber pathways and telecom components off-chip are common mode and do not affect the performance of the two-tone technique. Moreover, the compact size of the transceiver (5 mm × 20 mm) allows for uniform temperature control of the device. The PIC was co-packaged with a thermistor and a thermoelectric cooler (TEC) for temperature control to minimize the temperature coefficient of the transceiver, though temperature stabilization was not implemented to obtain the present results. One drawback of performing the optical interference entirely on chip is the 3 dB loss incurred by the 50/50 coupler that separates the local and received light.

The insertion loss from the terminal to the photodetector is measured to be 6 dB. This includes 3 dB from the use of a 50/50 coupler to separate the local and received light, 1 dB from the two chip-to-fiber interfaces, 0.5 dB for propagation and additional PIC splitter losses, and 1.5 dB from the telecom components used to wavelength-separate the light.

The CW lasers at each site are phase-stabilized to the frequency comb with a digital phase-locked loop filter and current modulation. Both CW lasers at a given site are phase locked to their nearest comb tooth with the same offset frequency. The difference in offset frequencies between the two sites determines the demodulation frequency in the digital signal processing and is chosen to be 10 MHz. Typical integrated residual phase noise for the two phase-locks is 0.75 rad in a 5 MHz bandwidth.

**Digital Signal Processing**
Extracting the one-way group delays is accomplished with digital signal processing on an FPGA at each site. The basic algorithm used to process the two beatnotes is depicted in Figure 8. The FPGA implements a numerically controlled oscillator (NCO) that is used to demodulate the optical beatnote between the local and remotely received lasers at the first optical frequency. The frequency of $NCO_1$ is constantly adjusted by a servo to keep it in quadrature with the incoming signal. By adjusting the frequency rather than the phase of $NCO_1$, there is no need to process phase wraps and the algorithm is effectively granted infinite phase range. After suitable adjustment of the phase-locked loop (PLL) parameters, the phase fluctuations of $NCO_1$ are the inverse of the relative phase fluctuations between the local and remote CW laser at the first optical frequency. Currently, the typical bandwidth for this servo is ~ 10 kHz due to decimation of the input signal and subsequent low pass filtering. We suspect that the limited servo gain creates a multiplicative error in the tracked phase. An improved servo scheme with larger bandwidth and gain is currently under development and has already demonstrated a ~2× decrease in the timing deviation (200 as @ 1 s at high received powers).

A frequency correction is added to $NCO_1$ to create $NCO_2$, which is then used to demodulate the optical beatnote between the local and remotely received lasers at the second optical frequency. The correction frequency is also constantly adjusted by a servo to null the phase fluctuations between $NCO_2$ and the second optical beatnote. The phase of $NCO_2$ then corresponds to the phase difference of the two remote CW lasers accumulated over the link, which is directly proportional to the one-way group delay. The integral of the correction frequency corresponds to the phase difference of the two remote CW lasers accumulated over the link, which is directly proportional to the one-way group delay. The FPGA runs at a clock rate of 200 MHz, and the one-way group delay is processed at a rate of 200 Hz.

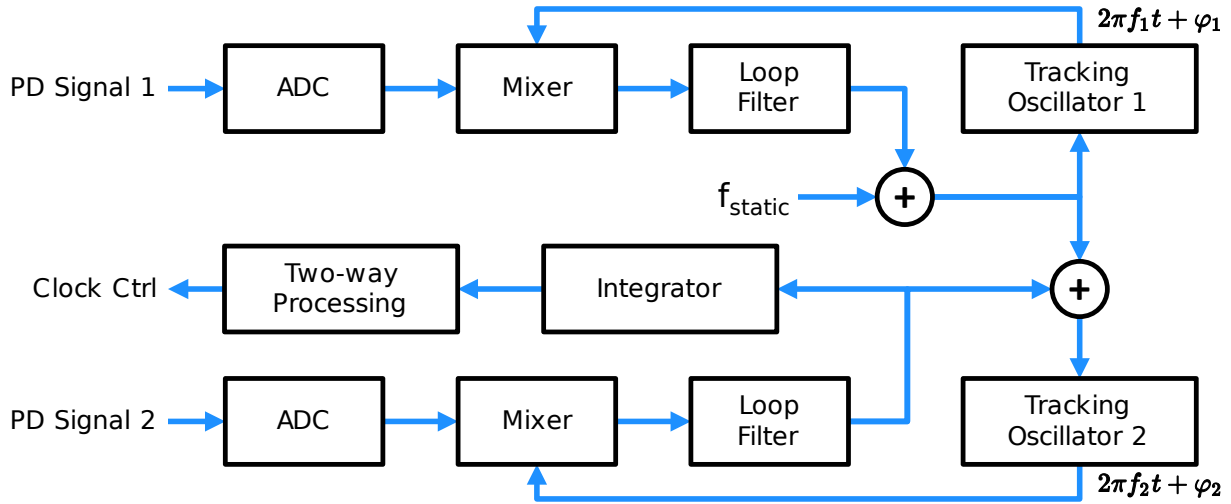

**Figure 8.** Signal processing block diagram for the two-tone technique. The heterodyne beatnote between the local and remote light is digitized for each of the two wavelengths. A numerical tracking oscillator is phase-locked to the first heterodyne beat, and the resultant phase of this tracking oscillator mirrors the optical phase variations $\varphi_1$. A second tracking oscillator, consisting of the sum between the first tracking oscillator and a correction signal, is then phase-locked to the second heterodyne beat, which likewise reproduces the optical phase variations $\varphi_2$. The correction signal is integrated to yield the phase difference between the two beatnotes $\varphi_2 - \varphi_1$, which is proportional to the one-way group delay.

## Acknowledgements

This research was developed with funding from DARPA contract HR001121C0175 (PRICELESS) and NAVAIR contract N6833523C0334 (PULSAR). The views, opinions and/ or findings expressed are those of the authors and should not be interpreted as representing the official views or policies of the Department of Defense or the US Government.

# Optical Two-Tone Time Transfer - Supplemental Material

Jonathan D. Roslund, Abijith S. Kowligy, Junichiro Fujita, Micah P. Ledbetter,
Akash V. Rakholia, Martin M Boyd, Jamil R Abo-Shaeer, Arman Cingöz

*Vector Atomic, Inc., Pleasanton, CA, USA*

(Dated: August 16, 2024)

The two-tone technique establishes an optical link between two remote sites, termed $A$ and $B$. At each site, two continuous wave lasers with indices 1 and 2 are phase-locked to a local frequency comb that is part of its optical reference. This process transfers timing noise-induced phase jitter of the frequency comb to the CW lasers. After stabilizing to the comb, each site's CW lasers are sent over the free-space (or fiber) link to the opposing site where they are coherently detected against the local CW lasers, which serve as local oscillators (LO).

## I. GENERAL DERIVATION OF TWO-TONE TIMING SIGNAL

The general electric field $E(t)$ for a CW laser at optical frequency $f$ is written as $E(t) = A \exp[\phi(\tau)] \exp[-2\pi i f t]$, where $A$ is its spectral amplitude and the spectral phase $\phi(\tau)$ depends upon the temporal variation $\tau$ of the comb pulse train to which the laser is locked. At Site A, the first CW laser is phase locked to the comb tooth at frequency $f_A^{(1)}$ with an offset frequency of $\Delta f_A^{(1)}$, and the second CW laser is phase locked to the comb tooth of frequency $f_A^{(2)}$ with an offset frequency of $\Delta f_A^{(2)}$. A similar scheme is employed at site $B$.

The two-tone technique employs coherent optical detection of the field sent from Site B to Site A, which is denoted as $E_{\text{sig},B \to A}^{(k)}$ where $k = 1, 2$ depending upon the CW laser. The light received from across the link is combined with the local field $E_{\text{LO},A}^{(k)}$ on a 50:50 beamsplitter located in the transceiver. The two output fields from this beamsplitter at Site A are written as:

$$E_{BS_1,A}^{(k)} = \left[ E_{\text{LO},A}^{(k)} + E_{\text{sig},B \to A}^{(k)} \right] / \sqrt{2}$$
$$E_{BS_2,A}^{(k)} = \left[ E_{\text{LO},A}^{(k)} - E_{\text{sig},B \to A}^{(k)} \right] / \sqrt{2}. \tag{1}$$

These two fields are detected with square-law detectors to generate photocurrent signals $i_{BS_j}^{(k)} = R \cdot E_{BS_j}^{(k)*} \cdot E_{BS_j}^{(k)}$ where $R = \Phi_{\text{qe}} \, e/\hbar\omega$ specifies the photodiode responsivity with $\Phi_{\text{qe}}$ representing the quantum efficiency. The photocurrent of the balanced detector formed by subtracting the photocurrents of the individual detectors is given by:

$$i_{\text{det},A}^{(k)} = i_{BS_1,A}^{(k)} - i_{BS_2,A}^{(k)}$$
$$= R \cdot \left( E_{\text{LO},A}^{(k)} E_{\text{sig},B \to A}^{(k)*} + E_{\text{LO},A}^{(k)*} E_{\text{sig},B \to A}^{(k)} \right) \tag{2}$$

To proceed, the LO and remotely received signal light at Site A are written in terms of their specific spectral amplitude and phase as:

$$E_{\text{LO},A}^{(k)} = A_A^{(k)} \exp\left[\phi_A^{(k)}(\tau)\right] \exp\left[-2\pi \left(f_A^{(k)} + \Delta f_A^{(k)}\right) \times t\right]$$
$$E_{\text{sig},B \to A}^{(k)} = \sqrt{T_{B \to A}^{(k)}} \, A_B^{(k)} \exp\left[\phi_B^{(k)}(\tau)\right] \exp\left[-2\pi \left(f_B^{(k)} + \Delta f_B^{(k)}\right) \times \left(t + n_{B \to A}^{(k)} L/c\right)\right], \tag{3}$$

where $T_{B \to A}^{(k)}$ is the optical link loss for CW laser $k$ in going from site $B$ to site $A$, $L$ is the length of the link, and $n_{B \to A}^{(k)}$ is the composite refractive index of the link for the Site B CW laser with index $k$. The substitution of these definitions into Eq. 2 allow the balanced photocurrent to be written in the form

$$i_{\text{det},A}^{(k)} = 2\,R \cdot \sqrt{T_{B \to A}^{(k)}} \, A_A^{(k)} \, A_B^{(k)} \times \cos\left[ \left(\phi_A^{(k)}(\tau) - \phi_B^{(k)}(\tau)\right) + 2\pi \left(\Delta f_A^{(k)} - \Delta f_B^{(k)} + \delta f^{(k)}\right) t \right. $$
$$\left. + 2\pi \left(f_B^{(k)} + \Delta f_B^{(k)}\right) n_{B \to A}^{(k)} L/c \right], \tag{4}$$

where $\delta f^{(k)} = f_A^{(k)} - f_B^{(k)}$ is the optical frequency difference between the Site A and Site B comb teeth to which the $k^{\text{th}}$ CW laser is locked.

This photocurrent is digitized and then IQ-demodulated at the user-defined offset frequency difference $\left(\Delta f_A^{(k)} - \Delta f_B^{(k)}\right)$ to extract the carrier phase $\varphi_A^{(i)}$, which is given as:

$$\varphi_A^{(i)} = \left[\phi_A^{(k)}(\tau) - \phi_B^{(k)}(\tau)\right] + 2\pi \delta f^{(k)} t + 2\pi \left(f_B^{(k)} + \Delta f_B^{(k)}\right) n_{B \to A}^{(k)} L/c. \quad (5)$$

A group phase $\varphi_{A,g}$ is then formed by subtracting the two carrier phases $\varphi_A^{(i)}$ at each site to yield:

$$\begin{aligned}\varphi_{A,g} &= \varphi_A^{(1)} - \varphi_A^{(2)} \\ &= \left[\phi_A^{(1)}(\tau) - \phi_A^{(2)}(\tau) - \phi_B^{(1)}(\tau) + \phi_B^{(2)}(\tau)\right] + 2\pi \left(\delta f^{(1)} - \delta f^{(2)}\right) t + \Phi_{B \to A}\end{aligned} \quad (6)$$

where $\Phi_{B \to A} = 2\pi \left[\left(f_B^{(1)} + \Delta f_B^{(1)}\right) n_{B \to A}^{(1)} - \left(f_B^{(2)} + \Delta f_B^{(2)}\right) n_{B \to A}^{(2)}\right] L/c$ is the optical group phase of the link experienced by the two CW lasers originating at Site B. Each term in Eq. 6 can be recast in a simpler form.

First, the time-varying optical phase $\phi_A^{(k)}(\tau)$ for each CW laser can be written in terms of the nearest comb tooth phase as

$$\begin{aligned}\phi_A^{(k)}(\tau) &\simeq \overline{\phi_A^{(k)}} + \phi_A|_{\omega_c} + d\phi_A(\omega)/d\omega|_{\omega_c} \cdot 2\pi (f_A^{(k)} - f_c) \\ &= \overline{\phi_A^{(k)}} + \phi_A|_{\omega_c} + \tau_A \cdot 2\pi (f_A^{(k)} - f_c)\end{aligned} \quad (7)$$

where $\overline{\phi_A^{(k)}}$ is the static phase for a given comb tooth (e.g., assuming that the pulse is not transform-limited), $\phi_A|_{\omega_c}$ is the comb phase at a convenient optical frequency $f_c = \omega_c/(2\pi)$, and $\tau_A$ is the temporal shift in the pulse envelope arrival time arising from an instability of the optical reference. With this definition, the first term of Eq. 6 is rewritten as:

$$\left[\phi_A^{(1)}(\tau) - \phi_A^{(2)}(\tau) - \phi_B^{(1)}(\tau) + \phi_B^{(2)}(\tau)\right] = \overline{\varphi_g} + 2\pi \left(\Delta f_A \tau_A - \Delta f_B \tau_B\right), \quad (8)$$

where $\overline{\varphi_g} = \overline{\phi_A^{(1)}} - \overline{\phi_A^{(2)}} - \overline{\phi_B^{(1)}} + \overline{\phi_B^{(2)}}$ is the static phase offset, $\Delta f_A = f_A^{(1)} - f_A^{(2)}$ is the CW laser frequency baseline at Site A, $\tau_A$ is the time instability at Site A, and similar definitions exist for the Site B terms. It is worth noting that the optical frequency $f_c$ does not appear in Eq. 8 with the natural assumption that the optical phase for the two CW lasers at a given site is Taylor-expanded about the same optical frequency. A natural choice for this expansion frequency is the arithmetic mean between the two CW laser frequencies at a given site. Different expansion frequencies between the two sites also does not affect Eq. 8 to first-order; however, higher-order corrections due to dispersion do appear and are considered in Sec. III.

The second term in Eq. 6, which gives an explicit time-dependence to the group phase and arises from a difference in the comb teeth frequencies between the two sites, is reformulated as:

$$\begin{aligned}\delta f^{(1)} - \delta f^{(2)} &= f_A^{(1)} - f_A^{(2)} - f_B^{(1)} + f_B^{(2)} \\ &= p \cdot f_{\text{rep},A} - q \cdot f_{\text{rep},B},\end{aligned} \quad (9)$$

where $f_{\text{rep}}$ is the repetition rate for the frequency comb at the corresponding site, and $p, q$ are integers representing the number of comb teeth separating the CW lasers at each site. Under typical circumstances, the CW laser pairs bridge the same number of comb teeth at each site (i.e., $p = q$), which then permits the additional simplification of $\delta f^{(1)} - \delta f^{(2)} = p \cdot \Delta f_{\text{rep}}$, where $\Delta f_{\text{rep}}$ is the difference in comb repetition rates between the two sites.

Third, the term in Eq. 6 describing the optical link phase $\Phi_{B \to A}$ is reworked in the following manner to yield:

$$\begin{aligned}
\Phi_{B \to A} &= 2\pi \left( f_B^{(1)} + \Delta f_B^{(1)} \right) n_{B \to A}^{(1)} \cdot L/c - 2\pi \left( f_B^{(2)} + \Delta f_B^{(2)} \right) n_{B \to A}^{(2)} \cdot L/c \\
&= \Phi_{B \to A}^{(1)} - \Phi_{B \to A}^{(2)} \\
&\simeq d\Phi_{B \to A}/d\omega \big|_{\omega_0} \cdot 2\pi \left( \Delta f_B + \Delta f_B^{(1)} - \Delta f_B^{(2)} \right) \\
&= d\Phi_{B \to A}/d\omega \big|_{\omega_0} \cdot 2\pi \Delta f_B \cdot \left( 1 + \frac{\Delta f_B^{(1)} - \Delta f_B^{(2)}}{\Delta f_B} \right) \\
&\simeq d\Phi_{B \to A}/d\omega \big|_{\omega_0} \cdot 2\pi \Delta f_B \\
&= T_{B \to A} \cdot 2\pi \, \Delta f,
\end{aligned} \quad (10)$$

where the time-of-flight (TOF) group-delay across the link $T_{B \to A}$ follows as the derivative of the link spectral phase with respect to angular frequency [1]. Additionally, the simplifying assumption has been made that the difference in the offset frequencies between the CW lasers at a given site is small compared to the separation of their optical frequencies, i.e., $\left( \Delta f_B^{(1)} - \Delta f_B^{(2)} \right)/\Delta f_B \ll 1$. Importantly, this assumption is made only for a simplified derivation of the key result. An accurate extraction of the TOF delay between the two sites is obtained by simplfying normalizing by the exact frequency baseline: $\Delta f_B \cdot \left( 1 + \frac{\Delta f_B^{(1)} - \Delta f_B^{(2)}}{\Delta f_B} \right)$. Finally, Eq. 6 is recast with the use of Eqs. 8, 9, and 10 to yield the one-way group phase:

$$\varphi_{A,g} = \overline{\varphi_g} + 2\pi \left( \Delta f_A \tau_A - \Delta f_B \tau_B \right) + 2\pi \, T_{B \to A} \cdot \Delta f_B + 2\pi \, p \cdot \Delta f_{\text{rep}} \, t. \quad (11)$$

For situations in which the repetition rates of the two combs are not equivalent (i.e., the timebases at sites A and B are not syntonized), this one-way group phase will accumulate linearly with time with a slope directly proportional to the frequency offset between the two combs. By observing the linear component of the group phase, a frequency adjustment may be performed on one of the frequency combs in order to syntonize the systems. Following such an adjustment, the one-way group phase simplifies to the form:

$$\varphi_{A,g} = \overline{\varphi_g} + 2\pi \Delta f \, \Delta \tau_{A,B} + 2\pi \, T_{B \to A} \cdot \Delta f, \quad (12)$$

where $\Delta \tau_{A,B} = \tau_A - \tau_B$ now represents the relative timebase instability between sites A and B and both frequency baselines for the CW lasers at each site are equivalent, i.e., $\Delta f_A = \Delta f_B \equiv \Delta f$. The one-way group phase is then normalized by this frequency baseline to yield the one-way group delay between the sites:

$$\begin{aligned}
\tau_{B \to A, g} &= \varphi_{A,g} / \left( 2\pi \Delta f \right) \\
&= \overline{\tau}_{B \to A, g} + \Delta \tau_{A,B} + T_{B \to A},
\end{aligned} \quad (13)$$

where $\overline{\tau}_{B \to A, g} = \overline{\varphi_g}/\left(2\pi \Delta f\right)$ represents a static time offset that may be neglected after designating an appropriate temporal reference plane at Site A to have a time offset of zero.

In order to separate the variable time-of-flight delay $T_{B \to A}$ from the timing signal $\Delta \tau_{A,B}$, the two-tone technique is operated in a two-way manner. As such, as anlogous detection and processing scheme is implemented at Site B, which yields its own one-way group delay given as:

$$\tau_{A \to B, g} = -\Delta \tau_{A,B} + T_{A \to B}, \quad (14)$$

where $\Delta \tau_{B,A} = -\Delta \tau_{A,B}$ since the magnitude of the timebase difference is independent of where it is measured, and the negative sign arises from shifting the frequency frame of reference from Site A to Site B. Again, the static time offset $\overline{\tau}_{A \to B, g}$ has been set to zero through appropriate definition of the temporal reference plane at Site B. The results from these two one-way measurements are exchanged between both sites by either an optical or microwave communication channel, which allows expressing the normalized difference of the two one-way delays as:

$$\left( \tau_{B \to A, g} - \tau_{A \to B, g} \right)/2 = \Delta \tau_{A,B} + \left( T_{B \to A} - T_{A \to B} \right)/2. \quad (15)$$

If the free-space link is assumed to be reciprocal (i.e., $T_{B \to A} = T_{A \to B}$), then the relative temporal jitter between the Site A and B timebases is finally written as:

$$\begin{aligned}\Delta \tau_{A,B} &= \left(\tau_{B \to A, g} - \tau_{A \to B, g}\right)/2 \\ &= \left(\varphi_{B \to A, g} - \varphi_{A \to B, g}\right)/4\pi \Delta f,\end{aligned} \quad (16)$$

which corresponds with the equation presented in the main text. The choice of the offset frequencies $\Delta f_{A,B}^{(k)}$ necessary to ensure link reciprocity will be discussed in Section III.

## II. DETECTION SENSITIVITY ANALYSIS

Next, a signal-to-noise analysis is performed for the two-tone technique in order to assess its sensitivity limit. This analysis proceeds by first considering the sensitivity limit for detecting a single CW laser at a given site and then proceeds to calculate the detection limit for the group delay.

### A. General Signal-to-Noise Determination

The measured photocurrent following coherent detection of a single CW tone at Site A is given by Eq. 4, which is recast as:

$$i_\text{det} = 2\, R\, \sqrt{P_\text{LO}\, P_\text{sig}} \times \cos\left[\phi_\text{LO} + \phi_\text{sig}\right], \quad (17)$$

where the optical powers for the local oscillator (LO) and detected beam are related to the field amplitudes by $P_\text{LO}^{(k)} = \left[A_A^{(k)}\right]^2$ and $P_\text{sig} = T_{B \to A}^{(k)} \cdot \left[A_B^{(k)}\right]^2$, respectively. The phase of the LO is proportional to the photodiode beatnote frequency and thus increases linearly with time as $\phi_\text{LO} = 2\pi \left(\Delta f_A^{(k)} - \Delta f_B^{(k)}\right) t$. Finally, the quantity of interest, which is the phase of the detected signal field is written as $\phi_\text{sig} = \left[\phi_A^{(k)}(\tau) - \phi_B^{(k)}(\tau)\right] + \Phi_{B \to A}^{(k)}$. Being a one-way signal, variations of this phase arise from both changes in the underlying timescales as well as fluctuations of the phase across the optical link. The peak value for the detected photocurrent $i_\text{peak}$ is then readily determined as $i_\text{peak} = 2\, R\, \sqrt{P_\text{LO}\, P_\text{sig}}$.

Any variation in the signal phase $\delta\phi_\text{sig}$ leads to a concomitant change in the measured photocurrent $\delta i_\text{det}$ given as

$$\begin{aligned}\delta i_\text{det} &= \left.\frac{\partial i_\text{det}}{\partial \phi_\text{sig}}\right|_{\phi_\text{LO}} \cdot \delta\phi_\text{sig} \\ &= -i_\text{peak} \cdot \sin(\phi_\text{LO}) \cdot \delta\phi_\text{sig} \\ \delta i_\text{det}^\text{rms} &= i_\text{peak} \cdot \sqrt{\langle \sin^2(\phi_\text{LO})\rangle} \cdot \delta\phi_\text{sig}^\text{rms}\end{aligned} \quad (18)$$

When the LO phase is $\phi_\text{LO} = \pi/2$, the phase quadrature of the detected field is probed and leads to the largest possible variation in the photocurrent, i.e., $\langle \sin^2(\phi_\text{LO})\rangle = 1$. Conversely, a LO phase of $\phi_\text{LO} = 0$ probes the amplitude quadrature of the signal field and obscures the phase variations. The minimum resolvable phase variation $\delta\phi_\text{sig, min}^\text{rms}$ is that which induces a photocurrent variation $\delta i_\text{det, min}^\text{rms}$ equal to the background photocurrent noise $\delta i_{\text{det},n}^\text{rms}$, i.e.,

$$\begin{aligned}\delta\phi_\text{sig, min}^\text{rms} &= \delta i_{\text{det},n}^\text{rms} / \left[i_\text{peak} \cdot \sqrt{\langle \sin^2(\phi_\text{LO})\rangle}\right] \\ &= 1/\left[\text{SNR} \cdot \sqrt{\langle \sin^2(\phi_\text{LO})\rangle}\right],\end{aligned} \quad (19)$$

where $\text{SNR} = i_\text{peak}/\delta i_{\text{det},n}^\text{rms}$ is the signal-to-noise ratio for the measurement. The RMS photocurrent noise $\delta i_{\text{det},n}^\text{rms}$ is comprised of two components. The first is power-dependent and arises from the quantum statistics of the detection process itself (i.e., the shot noise limit) while the second is power-independent and originates from non-common, residual noise on the phase locks of the CW laser to the frequency comb. These two types of photocurrent noise are uncorrelated and thus the total noise is constructed as their quadrature sum:

$$\left(\delta i_{\text{det},n}^{\text{rms}}\right)^2 = \left(\delta i_{\text{sn}}^{\text{rms}}\right)^2 + \left(\delta i_{\text{pll}}^{\text{rms}}\right)^2 \tag{20}$$

where $\delta i_{\text{sn}}^{\text{rms}}$ is the shot noise of the detection process and $\delta i_{\text{pll}}^{\text{rms}}$ is the classical noise arising from excess noise in the phase-locked loop.

In the situation that the LO power is significantly larger than the signal power, i.e., $P_{\text{LO}} \gg P_{\text{sig}}$, the signal beam contribution to the shot noise is insignificant compared to that arising from the LO. As such, the photocurrent variation arising from shot noise is written as [2]:

$$\begin{aligned}\left(\delta i_{\text{sn}}^{\text{rms}}\right)^2 &= 2\,e \cdot i_{\text{LO}} \cdot B \\ &= 2\,e\,R \cdot P_{\text{LO}} \cdot B,\end{aligned} \tag{21}$$

where the detection bandwidth $B$ is related to the sampling time interval $\Delta t$ through the relation $B = 1/2\Delta t$ and $R$ is once again the photodiode responsivity. The classical contribution to the photocurrent noise $\delta i_{\text{pll}}^{\text{rms}}$ originates from classical fluctuations of the signal phase $\Delta \phi_{\text{pll}}^{\text{rms}}$. These fluctuations arise from extraneous phase noise between the frequency comb and a single CW laser. The relation between these two quantities is defined by Eq. 18 to be:

$$\left(\delta i_{\text{pll}}^{\text{rms}}\right)^2 = i_{\text{peak}}^2 \cdot 2\left(\Delta \phi_{\text{pll}}^{\text{rms}}\right)^2. \tag{22}$$

This expression carries a multiplicative factor of 2 since each beatnote between two CW lasers convolves the excess phase noise present on the individual lasers with respect to their local frequency comb, and the assumption of approximately equal RMS phase noises has been made. The total photocurrent noise detected in the transceiver is then written by combining Eq. 21 and Eq. 22:

$$\left(\delta i_{\text{det},n}^{\text{rms}}\right)^2 = 2\,e\,R \cdot P_{\text{LO}} \cdot B + 2 \cdot i_{\text{peak}}^2 \cdot \left(\Delta \phi_{\text{pll}}^{\text{rms}}\right)^2. \tag{23}$$

The SNR is then readily found to be:

$$\begin{aligned}\text{SNR} &= i_{\text{peak}}/\delta i_{\text{det},n}^{\text{rms}} \\ &= 1/\sqrt{\frac{2\,e\,R \cdot P_{\text{LO}} \cdot B + 8\,R^2 \cdot P_{\text{LO}} \cdot P_{\text{sig}} \cdot \left(\Delta \phi_{\text{pll}}^{\text{rms}}\right)^2}{4\,R^2\,P_{\text{LO}}\,P_{\text{sig}}}} \\ &= \frac{1}{\sqrt{1/\left(4\,\Phi_{\text{qe}} \cdot N_{\text{sig}}\right) + 2 \cdot \left(\Delta \phi_{\text{pll}}^{\text{rms}}\right)^2}},\end{aligned} \tag{24}$$

where $\Phi_{\text{qe}}$ is again the quantum efficiency of the photodetector and $N_{\text{sig}}$ is the number of signal photons detected in the measurement time interval $\Delta t$. For the case in which the classical noise is negligible compared to the shot noise contribution, i.e., $\Delta \phi_{\text{pll}}^{\text{rms}} = 0$, the SNR assumes the simplified form of $\text{SNR} = 2\sqrt{\Phi_{\text{qe}} \cdot N_{\text{sig}}}$, which agrees with the result from a fully quantum treatment [2, 3]. The form of the SNR defined by Eq. 24 is then used to deduce the detection sensitivity for the two-tone measurement in both a one- and two-way configuration for the optical link.

### B. Two tone, Two-way Heterodyne Sensitivity

The detection limit for the two-tone, two-way method of time-transfer introduced in the main text follows directly from Eq. 19 and is written as:

$$\begin{aligned}\delta \phi_{\text{min}}^{\text{rms}} &= 1/\left[\text{SNR} \cdot \sqrt{\langle \sin^2(\phi_{\text{LO}})\rangle}\right] \\ &= \delta i_n^{\text{rms}}/i \times 1/\sqrt{\langle \sin^2(\phi_{\text{LO}})\rangle}.\end{aligned} \tag{25}$$

For situations in which the LO field is locked to the phase quadrature of the signal field (i.e., $\phi_{\rm LO} = \pi/2$, the denominator in Eq. 25 is given as $\langle \sin^2(\phi_{\rm LO})\rangle = 1$ and the detection sensitivity is maximized. However, for situations in which the LO is not phase-locked to the signal field, this term is of value less than unity. Typical demodulation frequencies in the digital processing unit for the present demonstration are in the 1-10 MHz range, which means that the LO scans through $\gtrsim 10^6$ cycles of the signal phase during a typical acquisition period of 1 sec. In this situation, we assume that $\langle \sin^2(\phi_{\rm LO})\rangle = 1/2$, which may be interpreted as a decrease in the effective integration time by a factor of 1/2.

An alternative, but equivalent interpretation for the decrease in the effective SNR is obtained in the frequency domain by considering the noise sidebands that contribute to the detected signal. Compared to homodyne detection (i.e., the LO and signal fields are of the same frequency), heterodyne detection detects vacuum fluctuations at the signal field frequency as well as its image below the LO [4, 5]. Thus, there is a $\sqrt{2}$ penalty for heterodyne readout since the LO folds the upper and lower signal vaccum sidebands together into the detection signal and they are uncorrelated. This 3dB penalty associated with heterodyne detection may be viewed as a consequence of the Heisenberg uncertainty relation since both the amplitude and phase quadratures of the signal field are permitted to be simultaneously detected [6].

After taking this penalty into account, the phase detection limit is recast as:

$$\begin{aligned}\delta\phi_{\rm min}^{\rm rms} &= \sqrt{2}/{\rm SNR} \\ &= \sqrt{2}\cdot \delta i_n^{\rm rms}/i.\end{aligned} \qquad (26)$$

The normalized composite photocurrent after combining the signal from all four detectors is written as:

$$\begin{aligned}i &= \left(i_A^{(1)} - i_A^{(2)} - i_B^{(1)} + i_B^{(2)}\right)/2 \\ &= \left[i_{A,{\rm peak}}^{(1)}\cdot\sin\left(\phi_A^{(1)}\right) - i_{A,{\rm peak}}^{(2)}\cdot\sin\left(\phi_A^{(2)}\right) - i_{B,{\rm peak}}^{(1)}\cdot\sin\left(\phi_B^{(1)}\right) + i_{B,{\rm peak}}^{(2)}\cdot\sin\left(\phi_B^{(2)}\right)\right]/2 \\ &\simeq \left[i_{A,{\rm peak}}^{(1)}\cdot\phi_A^{(1)} - i_{A,{\rm peak}}^{(2)}\cdot\phi_A^{(2)} - i_{B,{\rm peak}}^{(1)}\cdot\phi_B^{(1)} + i_{B,{\rm peak}}^{(2)}\cdot\phi_B^{(2)}\right]/2,\end{aligned} \qquad (27)$$

where we have made the small angle approximation for the two phases under study (i.e., $\sin(\phi)\simeq\phi$). It is important to note that use of the small angle approximation is only for the present signal-to-noise analysis since the IQ demodulation scheme is capable of extracting a phase of arbitrary magnitude. Next, we make the reasonable assumption that the same number of signal photons is measured at all four photodetectors (i.e., $P_{{\rm sig},A}^{(1)} = P_{{\rm sig},A}^{(2)} = P_{{\rm sig},B}^{(1)} = P_{{\rm sig},B}^{(2)}$), which allows writing $i_{A,B,{\rm peak}}^{(k)} = i_{A,{\rm peak}}^{(1)}$. The group phase at a single site is formed by subtracting the two carrier phases, i.e., $\phi_{A,g} = \phi_A^{(1)} - \phi_A^{(2)}$ with an analogous definition at Site B. Subtraction of the two group phases $\phi_{A,g}$ and $\phi_{B,g}$ removes the common-mode term between them (i.e., the TOF signal as in Eq. 15) and leaves the relative jitter from the two timebases $\Delta\tau_{A,B}$. Since $\Delta\tau_{A,B} = -\Delta\tau_{B,A}$ by definition, the sign relationship between the two carrier phases is taken as $\phi_{A,g} = -\phi_{B,g}$. The composite photocurrent then simplifies to

$$\begin{aligned}i &\simeq i_{A,{\rm peak}}^{(1)}\cdot\left[\phi_A^{(1)} - \phi_A^{(2)} - \phi_B^{(1)} + \phi_B^{(2)}\right]/2 \\ &\simeq i_{A,{\rm peak}}^{(1)}\cdot\left[\phi_{A,g} - \phi_{B,g}\right]/2 \\ &\simeq i_{A,{\rm peak}}^{(1)}\cdot\phi_{A,g}.\end{aligned} \qquad (28)$$

Next, the RMS fluctuations for this composite photocurrent is given as:

$$\langle(\delta i_n)^2\rangle = \left[\langle\left(\delta i_A^{(1)}\right)^2\rangle + \langle\left(\delta i_A^{(2)}\right)^2\rangle + \langle\left(\delta i_B^{(1)}\right)^2\rangle + \langle\left(\delta i_B^{(2)}\right)^2\rangle - 2\cdot\langle\delta i_A^{(1)}\cdot\delta i_B^{(1)}\rangle - 2\cdot\langle\delta i_A^{(2)}\cdot\delta i_B^{(2)}\rangle\right]/4, \qquad (29)$$

where the correlation between two photocurrents at a given site is zero (i.e., $\langle\delta i_A^{(1)}\cdot\delta i_A^{(2)}\rangle = \langle\delta i_B^{(1)}\cdot\delta i_B^{(2)}\rangle = 0$) as the shot noise between separate detectors is uncorrelated and the residual PLL noise is also uncorrelated since it arises from phase-locking to separate lasers. We then proceed to make the reasonable assumption that the RMS photocurrent noise at the four individual detectors is equivalent (i.e., the residual PLL noise for all four CW locks is approximately the same, $\langle\left(\delta i_A^{(1)}\right)^2\rangle = \langle\left(\delta i_A^{(2)}\right)^2\rangle = \langle\left(\delta i_B^{(1)}\right)^2\rangle = \langle\left(\delta i_B^{(2)}\right)^2\rangle$). Any correlation between the photocurrent noises at separated sites must be classical in nature since the shot noise between individual detectors is uncorrelated. As such, we assume that such a classical correlation between the two sites does not depend upon the CW index $k$, i.e., $\langle \delta i_A^{(1)} \cdot \delta i_B^{(1)} \rangle = \langle \delta i_A^{(2)} \cdot \delta i_B^{(2)} \rangle$ (the origin for such a classical correlation will be discussed henceforth, which justifies this assumption). The RMS fluctuations for the composite photocurrent then simplifies to the form:

$$\langle (\delta i_n)^2 \rangle \simeq \langle \left(\delta i_A^{(k)}\right)^2 \rangle - \langle \delta i_A^{(k)} \cdot \delta i_B^{(k)} \rangle \tag{30}$$

The shot-noise component of the photocurrent fluctuations of course does not exhibit correlation between the two sites. However, classical correlations can exist and the covariance $\langle \delta i_A^{(1)} \cdot \delta i_B^{(1)} \rangle$ is not strictly zero. For links of zero length, the classical PLL noise between sites detecting the same CW beatnote is perfectly anticorrelated. This is readily understood by realizing that both sites are measuring a beatnote between the same two lasers. The noise recorded at Site A is from its frequency reference frame, i.e., $\delta f_{A,B}^{(k)} = f_A^{(k)} - f_B^{(k)}$ while the same fluctuations measured at Site B are from its own frame, i.e., $\delta f_{B,A}^{(k)} = f_B^{(k)} - f_A^{(k)}$. Thus, we see that $\delta f_{A,B}^{(k)} = -\delta f_{B,A}^{(k)}$. In the limit of an infinitely long link, however, the classical correlation is zero, which is explained by the finite speed of light. If the travel time between the two sites exceeds the coherence time for the stabilized CW laser, the beatnotes from different sites are independent and the correlation goes to zero. This effect is entirely analogous to the measurement of laser linewidths by heterodyning a laser against a time-delayed version of itself [7]. Notably, the same argument holds for the sum of the photocurrents, which reveals the TOF for the link. Due to the anti-correlation of the classical PLL noise for links shorter than the CW laser coherence length, the sum removes the classical contribution completely and the resulting TOF signal is purely shot noise limited (e.g., see Fig. 4b in the main text). After accounting for this classical correlation, the composite photocurrent RMS fluctuations across the link are written as:

$$\langle (\delta i_n)^2 \rangle = (\delta i_{\rm sn}^{\rm rms})^2 + \left(\delta i_{\rm pll}^{\rm rms}\right)^2 \cdot (1+g) \tag{31}$$

where $g=0$ for an uncorrelated link (i.e., the transit time exceeds the coherence time for the stabilized CW laser) and $g=1$ for a fully correlated link (link transit time is much shorter than the coherence time for the CW lasers). Following the use of Eq. 21 and Eq. 22 to describe the individual terms $(\delta i_{\rm sn}^{\rm rms})^2$ and $\left(\delta i_{\rm pll}^{\rm rms}\right)^2$, respectively, the RMS fluctuations for the composite photocurrent simplifies to:

$$\delta i_n^{\rm rms} = \sqrt{2\,e\,R \cdot P_{\rm LO} \cdot B + 2 \cdot \left(i_{A,\rm peak}^{(k)}\right)^2 \cdot \left(\Delta \phi_{\rm pll}^{\rm rms}\right)^2 \cdot (1+g)} \tag{32}$$

This expression is combined with Eq. 28 for describing the peak composite photocurrent and substituted into Eq. 25 to yield the measurement SNR:

$$\begin{aligned}
{\rm SNR} &= 1/\sqrt{1/\left(4\,\Phi_{\rm qe} \cdot N_A^{(k)}\right) + 2 \cdot (1+g) \cdot \left(\Delta \phi_{\rm pll}^{\rm rms}\right)^2} \\
&= 1/\sqrt{1/\left(\Phi_{\rm qe} \cdot N_{A+B}\right) + 2 \cdot (1+g) \cdot \left(\Delta \phi_{\rm pll}^{\rm rms}\right)^2}
\end{aligned} \tag{33}$$

where $N_{A+B} = N_A^{(1)} + N_A^{(2)} + N_B^{(1)} + N_B^{(2)} \simeq 4 \cdot N_A^{(k)}$ is the total number of measured photons across all four photodetectors in the sampling time interval $\Delta t$, and the photons impinging upon a single photodetector is $N_A^{(k)} = P_{\rm sig, A}^{(k)}/(h\nu) \cdot \Delta t$. Thus, the SNR associated with two-tone, two-way time transfer is determined by two parts. The first term is power-dependent and exhibits the $1/\sqrt{N}$ form that is typical for shot-noise limited detection. The second term is power-independent and arises from the residual phase noise between the CW laser and the comb tooth to which it is locked.

$$\begin{aligned}
\delta \phi_{\rm min}^{\rm rms} &= \sqrt{2/\left(\Phi_{\rm qe} \cdot N_{A+B}\right) + 4 \cdot (1+g) \cdot \left(\Delta \phi_{\rm pll}^{\rm rms}\right)^2} \\
&= \sqrt{2/\left(\Phi_{\rm qe} \cdot N_{A+B}\right) + \kappa \cdot \left(\Delta \phi_{\rm pll}^{\rm rms}\right)^2}
\end{aligned} \tag{34}$$

where $\langle\sin^2(\phi_{\text{LO}})\rangle = 1/2$ has been utilized and $\kappa$ is defined as $\kappa = 4 \cdot (1+g)$. Thus, $\kappa = 4$ for a long link (i.e., the transit time exceeds the coherence time for the stabilized CW laser) and $\kappa = 8$ for a short link.

It is interesting to compare the phase measurement sensitivity for the group delay given by Eq. 34 to the standard-quantum limit for an interferometer $\delta\phi_{\text{SQL}}^{\text{rms}} = 1/\left(2\sqrt{N}\right)$ given a fixed number of photons [8–10]. In the limit where shot-noise dominates the sensitivity (i.e., the low photon limit), the ratio of these sensitivities is given as $\left[\delta\phi_{\text{min}}^{\text{rms}}/\delta\phi_{\text{SQL}}^{\text{rms}}\right] = \left(\sqrt{2}\right)^3$. The first factor is $\sqrt{2}$ is straightforward to understand and results from the fact that the LO is not locked onto the quadrature of the signal field but rather rapidly scans through it as discussed above. A second factor of $\sqrt{2}$ originates from the fact that knowledge of the group phase comes at the expense of dividing the total number of photons detected at a given site between two different carrier phase measurements. The final factor of $\sqrt{2}$ arises from the combination of information from these two uncorrelated detectors at a given site (i.e., the photocurrent noise for the two detectors in a single transceiver is summed in quadrature).

Finally, the group phase sensitivity is converted to the relevant timing sensitivity for the two-way, two-tone technique after multiplying by the discriminant $T/2\pi$ to yield:

$$\begin{aligned}\tau_{\text{g}}^{\text{rms}} &= \frac{T}{2\pi} \cdot \delta\phi_{\text{min}}^{\text{rms}} \\ &= \frac{T}{2\pi} \cdot \sqrt{2/\left(\Phi_{\text{qe}} \cdot N_{\text{A+B}}\right) + \kappa \cdot \left(\Delta\phi_{\text{pll}}^{\text{rms}}\right)^2},\end{aligned} \quad (35)$$

which corresponds with the sensitivity equation presented in the main text.

## III. EFFECT OF LINK DISPERSION

This section discusses selection of the various offset lock frequencies for the CW lasers in order to ensure reciprocity as well as the degradation of the link reciprocity due to material dispersion. To start, the spectral phase for the optical link is written as:

$$\phi(\omega) = k(\omega) \cdot L, \quad (36)$$

where $L$ is the length of the common-mode link and $k(\omega) = (\omega/c) \cdot n(\omega)$ is the frequency-dependent propagation constant for the link in which $n(\omega)$ is the refractive index. The angular frequencies for the two CW lasers at each site are given as $\omega_{A,B}^{(1)} = \omega_c + \Delta\omega/2 + \Delta_{A,B}^{(1)}$ and $\omega_{A,B}^{(2)} = \omega_c - \Delta\omega/2 + \Delta_{A,B}^{(1)}$, where $\Delta\omega$ is the angular frequency separation between the two CW lasers (e.g., $2\pi\cdot$ 200GHz frequency spacing for lasers separated by two ITU channels) and $\Delta_{A,B}^{(k)}$ is the angular frequency offset between the CW laser with index $k$ and the nearest comb tooth. The frequency $\omega_c$ serves as a virtual bookkeeping frequency that is common to both sites A and B and about which all of the optical phases are Taylor-expanded. While used for bookkeeping reasons, the center frequency $\omega_c$ does represent a measurable quantity and is defined as the arithmetic mean of the two comb teeth to which the CW lasers are locked: $2\pi\,\omega_c = (p+q)/2 \cdot f_{\text{rep}} + f_{\text{ceo}}$ where $p$ and $q$ are the relevant mode numbers.

With these definitions, the optical phase accumulated by CW laser 1 across the link from Site B to A is written as:

$$\phi_{B \to A}^{(1)} = k(\omega_c + \Delta\omega/2 + \Delta_B^{(1)}) \cdot L, \quad (37)$$

and a similar equation exists for the phase accumulated by CW laser 2. This optical phase is Taylor-expanded to third-order about the frequency $\omega_c$ to yield

$$\phi_{B \to A}^{(1)} = \phi(\omega_c) + \phi^{(1)} \cdot (\Delta\omega/2 + \Delta_B^{(1)}) + \phi^{(2)}/2 \cdot (\Delta\omega/2 + \Delta_B^{(1)})^2 + \phi^{(3)}/6 \cdot (\Delta\omega/2 + \Delta_B^{(1)})^3 + \ldots, \quad (38)$$

where $\phi^{(n)} = d^n\phi(\omega - \omega_c)/d\omega^2 \big|_{\omega=\omega_c}$ is the $n^{\text{th}}$-order derivative of the spectral phase evaluated at the frequency $\omega_c$. Similarly, the spectral phase accumulated by CW laser 2 across the link from Site B to A is written as:

$$\phi_{B \to A}^{(2)} = \phi(\omega_0) + \phi^{(1)} \cdot (-\Delta\omega/2 + \Delta_B^{(2)}) + \phi^{(2)}/2 \cdot (-\Delta\omega/2 + \Delta_B^{(2)})^2 + \phi^{(3)}/6 \cdot (-\Delta\omega/2 + \Delta_{(B)}^1)^3 + \ldots. \quad (39)$$

The group phase is formed by subtracting these two carrier phases, which is written as:

$$\Delta\phi_{B\to A} = \phi^{(1)}_{B\to A} - \phi^{(2)}_{B\to A}$$
$$= \phi^{(1)}\left(\Delta\omega + \Delta^{(1)}_B - \Delta^{(2)}_B\right) + \phi^{(2)}/2\left[\left(\Delta^{(1)}_B\right)^2 - \left(\Delta^{(2)}_B\right)^2 + \Delta\omega\left(\Delta^{(1)}_B + \Delta^{(2)}_B\right)\right] + \quad (40)$$
$$+ \phi^{(3)}/6\left[\Delta\omega^3/4 + 3\left(\Delta^{(1)}_B - \Delta^{(2)}_B\right)\Delta\omega^2/4 + 3\left((\Delta^{(1)}_B)^2 + (\Delta^{(2)}_B)^2\right)\Delta\omega/2 + \left(\Delta^{(1)}_B)^3 - \Delta^{(2)}_B)^3\right)\right]\ldots$$

Likewise, the phase difference $\Delta\phi_{A\to B}$ is written as:

$$\Delta\phi_{A\to B} = \phi^{(1)}_{A\to B} - \phi^{(2)}_{A\to B}$$
$$= \phi^{(1)}\left(\Delta\omega + \Delta^{(1)}_A - \Delta^{(2)}_A\right) + \phi^{(2)}/2\left[\left(\Delta^{(1)}_A\right)^2 - \left(\Delta^{(2)}_A\right)^2 + \Delta\omega\left(\Delta^{(1)}_A + \Delta^{(2)}_A\right)\right] \quad (41)$$
$$+ \phi^{(3)}/6\left[\Delta\omega^3/4 + 3\left(\Delta^{(1)}_A - \Delta^{(2)}_A\right)\Delta\omega^2/4 + 3\left((\Delta^{(1)}_A)^2 + (\Delta^{(2)}_A)^2\right)\Delta\omega/2 + \left(\Delta^{(1)}_A)^3 - \Delta^{(2)}_A)^3\right)\right]\ldots$$

The two-way timing phase $\Delta\phi^{\text{nr}}_{A,B}$ arising from the link itself is given as the normalized difference between these two one-way group phases:

$$\Delta\phi_{A,B} = \frac{1}{2}\left(\phi_{B\to A} - \phi_{A\to B}\right). \quad (42)$$

A fully reciprocal optical link would exhibit a two-way phase of zero, i.e., $\Delta\phi_{A,B} = 0$, and any non-zero phase is indistinguishable from a relative offset in the two timebases, thus degrading the accuracy of the measurement. Hence, the offset phases for the method must be chosen to work in concert with the relevant material dispersion in order to minimize any non-reciprocal residuals of the two-way phase. In general, any index designating the originating site (i.e., A or B) in Eq. 40 or Eq. 41 indicates that the link is unique to the particular CW lasers at a site. Therefore, in order to have a reciprocal link, these two phase differentials should not depend upon the index designating the site of origin. All site-dependent labels in these expressions are contained in the individual offset lock frequencies $\Delta^{(k)}_{A,B}$. As such, the link reciprocity is determined by judicious selection of these offset frequencies. There are two natural selections for these offset lock frequencies that minimize link non-reciprocity.

### A. Offset Frequency Choice #1

The site dependence for the leading terms in Eq. 40 and Eq. 41 is removed if both offset frequencies at a given site are equal, which is the approach adopted in the present work. Such a selection allows rewriting the individual offsets as $\Delta^{(1)}_A = \Delta^{(2)}_A \equiv \Delta_A$ and $\Delta^{(1)}_B = \Delta^{(2)}_B \equiv \Delta_B$. Given this selection for the offset frequencies, the timing phase $\Delta\phi_{A,B}$ is equivalent to:

$$\Delta\phi^{\text{nr}}_{A,B} \simeq k^{(2)}\, L\, \Delta\omega\, (\Delta_B - \Delta_A)/2. \quad (43)$$

This phase is converted to a non-reciprocal timing deviation after normalizing by the frequency separation baseline $\Delta\omega$ to yield:

$$\Delta t_{A,B} = \simeq k^{(2)}\, L\, (\Delta_B - \Delta_A)/2. \quad (44)$$

Thus, the error is directly proportional to the product of the link group velocity dispersion and the beatnote frequency detected at each site. As a quantitative example, the group velocity dispersion $k^{(2)}$ of air at 1560nm is $k^{(2)} = 1.06 \cdot 10^{-29}\,\text{s}^2/\text{m}$ [11]. A typical beatnote detection frequency is $\left(\Delta^B - \Delta^A\right) = 2\pi \cdot 10\,\text{MHz}$. The non-reciprocal timing error is then found to be $\Delta t_{A,B} = 3.3 \cdot 10^{-19}\,\text{s/km}$, and a timing error of 1 fs is developed after a propagation distance of $\sim 3000\,\text{km}$.

Further insight into this result may be had through the use of a simplified model for the two-tone technique. If we consider the superposition of two co-propagating CW fields of amplitude $A_0$ and with angular frequencies $\omega_\alpha$ and $\omega_\beta$, the resulting field is the product of a carrier and a temporal envelope, i.e.,

$$\begin{aligned} E_{\text{total}} &= A_0 \exp\left[-i\omega_\alpha t\right] + A_0 \exp\left[-i\omega_\beta t\right] \\ &= 2A_0 \exp\left[-i\left(\omega_\alpha + \omega_\beta\right)t/2\right] \times \cos\left[\left(\omega_\alpha - \omega_\beta\right)t/2\right] \\ &= 2A_0 \exp\left[-i\omega_\gamma t\right] \times \cos\left[\left(\omega_\alpha - \omega_\beta\right)t/2\right], \end{aligned} \quad (45)$$

where the carrier frequency $\omega_\gamma$ is the average of the two constituent angular frequencies, i.e., $\omega_\gamma = (\omega_\alpha + \omega_\beta)/2$. The envelope propagates at the group velocity and exhibits a period $T$ given as $T = 2\pi/(\omega_\alpha - \omega_\beta)$. The temporal delay experience by this envelope in transiting a dispersive link is written as $\tau(\omega_\gamma - \omega_c) = \tau_c + k^{(2)}\, L\, (\omega_\gamma - \omega_c)$, where $\tau_c$ is the group delay at a convenient bookkeeping frequency of $\omega_c$.

This model is then mapped onto the two-tone approach by realizing that the carrier frequencies $\omega_\gamma$ for the CW laser pairs at sites A and B are different. At Site A, the carrier frequency for the two CW fields is displaced from the bookkeeping frequency by the offset lock frequency, i.e., $\omega_{\gamma,A} = \omega_c + \Delta_A$. With this definition, its transit time across the link is written as:

$$\tau_{A \to B} = \tau_c + k^{(2)}\, L\, \Delta_A. \quad (46)$$

The carrier frequency for the two fields at Site B is displaced from the same bookkeeping frequency by its offset lock frequency, i.e., $\omega_{\gamma,B} = \omega_c + \Delta_B$, and the corresponding transit time across the link is

$$\tau_{B \to A} = \tau_c + k^{(2)}\, L\, \Delta_B. \quad (47)$$

The non-reciprocal timing between these two one-way transit times is then

$$\begin{aligned} \Delta t_{A,B} &= \left(\tau_{A \to B} - \tau_{B \to A}\right)/2 \\ &= k^{(2)}\, L\, (\Delta_A - \Delta_B)/2, \end{aligned} \quad (48)$$

which agrees with Eq. 44. Hence, the choice of different offset frequencies at the two sites disrupts the symmetry around the virtual carrier frequency $\omega_c$ common to both sites, and it is the difference in group delays between these two effective carriers that results in a non-reciprocal timing error.

### B.  Offset Frequency Choice #2

An alternative choice for the offset frequencies is given as $\Delta_A^{(1)} + \Delta_A^{(2)} = \Delta_B^{(1)} + \Delta_B^{(1)} = 0$. This corresponds to the two offset frequencies at each site having equal magnitudes but opposite signs (this situation is straightforward to achieve by simply locking the two CW lasers at each site to the relevant comb teeth with the same offset frequency magnitude but opposite signs). With this convention, the composite phase differential at Site A and B is found to be:

$$\begin{aligned} \Delta\phi_{B \to A} &= \phi^{(1)}\left(\Delta\omega + 2\,\Delta_B\right) + \phi^{(3)} \cdot \left[\Delta\omega^3/24 + \Delta\omega^2 \cdot \Delta_B/4 + \Delta\omega \cdot (\Delta_B)^2/2 + (\Delta_B)^3/3\right]\ldots \\ \Delta\phi_{A \to B} &= \phi^{(1)}\left(\Delta\omega + 2\,\Delta_A\right) + \phi^{(3)} \cdot \left[\Delta\omega^3/24 + \Delta\omega^2 \cdot \Delta_A/4 + \Delta\omega \cdot (\Delta_A)^2/2 + (\Delta_A)^3/3\right]\ldots \end{aligned} \quad (49)$$

Of particular interest is the fact that the quadratic term is absent in these expressions. For Offset Choice #1, the leading error term results from the fact that the offset frequencies displace the arithmetic mean of the two CW laser frequencies from the common frequency $\omega_c$. Offset Choice #2 purposely selects the offset frequencies at each site so as to maintain the same CW laser arithmetic mean between the two sites. In general, the group phase difference between two frequencies oppositely spaced about $\omega_c$ is sensitive to only odd-order terms in the Taylor expansion about $\omega_c$. Thus, this selection of offset frequencies preserves the symmetry and eliminates the second-order dispersive term.

One complication associated with this selection is that the frequency baseline between the two sites is now different, i.e., it is $\Delta\omega + 2\,\Delta_A$ for one site and $\Delta\omega + 2\,\Delta_B$ for the other. These unequal baselines would render the link non-reciprocal if not accounted for before exchanging phases between the two sites. However, the factor $\Delta\omega + 2\Delta_{A,B}$ is entirely determined by experimental parameters and is therefore in principle a known quantity.

For convenience, this frequency baseline is rewritten as:

$$\Delta\omega + 2\,\Delta_A = 2\pi \cdot p \cdot f_{\text{rep}} + 2\,\Delta_A, \quad (50)$$

where $f_{\text{rep}}$ is the known repetition rate of the frequency comb and $p$ is an integer describing the number of comb teeth separating the nascent frequency separation of the two CW lasers. Thus, this factor may be very precisely characterized and utilized to properly normalize the individual CW frequency baselines at sites A and B. In order to do so, the two sites now form a weighted superposition of their individually measured one-way time of flights to yield:

$$\Delta t_{A,B}^{\text{error}} = \left(\tau_{B \to A} - \gamma \cdot \tau_{A \to B}\right)/2 = \frac{1}{2} \cdot \phi^{(1)} \cdot \left[(1 + 2\Delta_B/\Delta\omega) - \gamma \cdot (1 + 2\Delta_A/\Delta\omega)\right] + \ldots \quad (51)$$

If the weighting factor is taken to be:

$$\begin{aligned}\gamma &= \frac{1 + 2\Delta^B/\Delta\omega}{1 + 2\Delta^A/\Delta\omega} \\ &\simeq 1 + \frac{2}{\Delta\omega} \cdot (\Delta^B - \Delta^A),\end{aligned} \quad (52)$$

then the first-order error to the timing signal disappears, and the residual error is given as:

$$\begin{aligned}\tau_{\text{error}} &\simeq \phi^{(3)}/2 \cdot \left[\Delta\omega^2/24 \cdot (1 - \gamma) + \Delta\omega/4 \cdot \left(\Delta^B - \gamma \Delta^A\right) + \ldots\right] \\ &\simeq k^{(3)}/12 \cdot \Delta\omega \cdot (\Delta^B - \Delta^A) \cdot L\end{aligned} \quad (53)$$

As a quantitative example, the third-order dispersion $k^{(3)}$ of air at 1560nm is $k^{(3)} = 9 \cdot 10^{-45}\,\text{s}^3/\text{m}$. If we take the coarse separation between the two CW lasers to be $\Delta\omega = 2\pi \cdot 200\,\text{GHz}$ and a typical beatnote detection frequency of $\left(\Delta^B - \Delta^A\right) = 2\pi \cdot 10\,\text{MHz}$, the non-reciprocal timing error over the link is $5.9 \cdot 10^{-23}\,\text{s}$ / km, which is $\sim 5000$ times smaller than the link error obtained using Offset Choice #1. The link distance at which the timing error is 1fs is $\sim 17 \cdot 10^6$ km.

Finally, although not outlined here, a third method for implementing the CW time transfer approach employs three CW lasers at each site. The availability of three lasers allows the elimination of dispersive effects up to third order and is valuable in situations where symmetry can not be exploited (e.g., the coupling of Doppler effects with dispersion associated with moving targets [12]). This is achieved with a direct projective measurement of both the group delay and its dispersion.

## IV. DEMONSTRATION OF SYNTONIZATION BETWEEN TWO IODINE CLOCKS

The comb repetition rates for two iodine clocks participating in the comparison measurement are initially syntonized by performing a microwave phase comparison (using a 53100A) and then appropriately adjusting the comb offset frequencies to null any linear phase accumulation. The optical comb tooth number (i.e., the comb tooth number that is phase-locked to the optically-stabilized laser) is determined by both systems, which relates the change in the comb offset lock frequencies necessary to induce a target offset in the repetition rates. These offset frequencies are then changed to induce a coarse frequency offset of $5 \cdot 10^{-11}$ between the two systems. Measurement of the time offset between the two systems is performed across the free-space test link with the two-tone technique, and the observed phase accumulation is seen in Fig. 1a. The phase is seen to accumulate linearly in time with a slope given by 50 ps/s, which corresponds with the expected offset of $5 \cdot 10^{-11}$. A fine frequency offset of value $7 \cdot 10^{-15}$ is created next, and the timing signal observed over the link is shown in Fig. 1b. In this case, the temporal fluctuations between the clocks is of similar magnitude to the linear phase accumulation arising from the intentional frequency offset. However, a linear fit yields a slope of 7 fs/s, which agrees with the expected offset. Thus, the two-tone technique provides a straightforward way to syntonize multiple optical references.

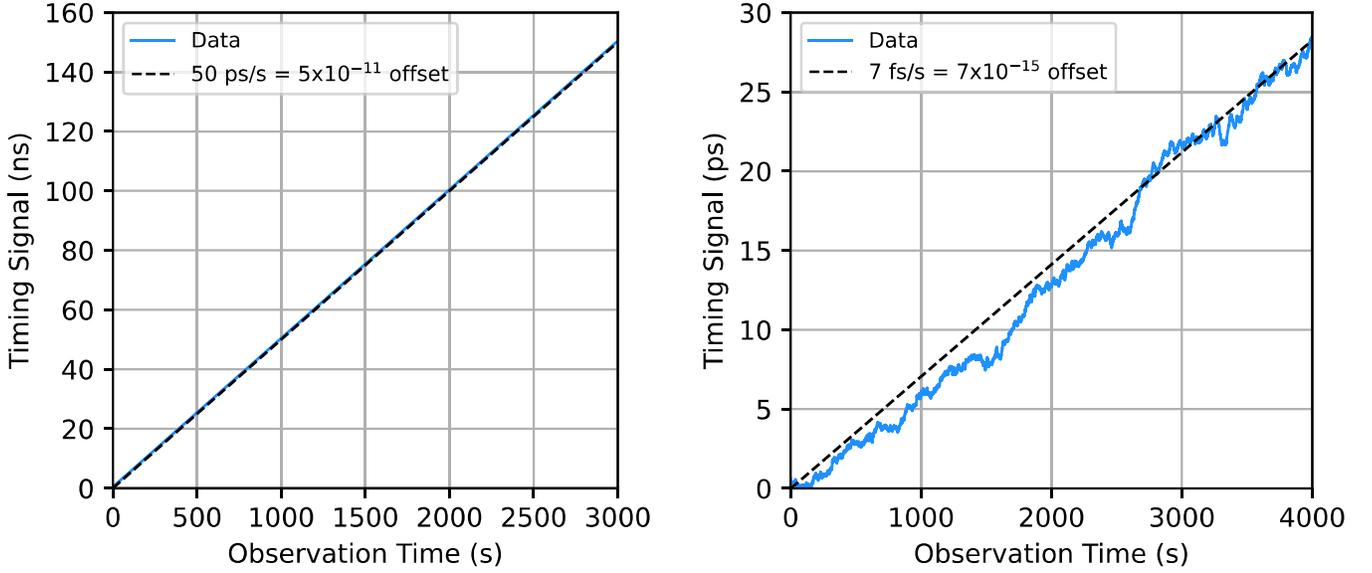

FIG. 1. Syntonization of iodine clocks with the two-tone method. The frequency comb offset lock frequencies are purposefully changed with knowledge of the comb tooth number to induce a desired frequency offset between the two systems. A) The frequency offset is altered to create a $5 \cdot 10^{-11}$ offset between the two systems. A linear phase accumulation of 50 ps/s is measured between the two systems with the two-tone technique. B) The frequency offset is altered to create a $7 \cdot 10^{-15}$ offset between the two systems, and a corresponding linear phase of slope 7 fs/s is measured.